\documentclass[aps,pre,preprint,groupedaddress]{revtex4-1}
\usepackage{amsmath}
\usepackage{graphicx}

\begin{document}


\title{Modelling stochastic resonance in humans: the influence of lapse rate}


\author{Jeroen J.A. van Boxtel}
\email[]{j.j.a.vanboxtel@gmail.com}
\homepage[]{http://users.monash.edu/~jeroenv}
\thanks{}
\affiliation{School of Psychological Sciences and Monash Institute of Cognitive and Clinical Neurosciences, Monash University, Clayton 3800 Vic, Australia}
\date{\today}

\begin{abstract}
Adding noise to a sensory signal generally decreases human performance. However noise can  improve performance too, due to a process called stochastic resonance (SR). This paradoxical effect may be exploited in psychophysical experiments, to provide additional insights into how the sensory system deals with noise. Here, I develop a model for stochastic resonance to study the influence of noise on human perception, in which the biological parameter of `lapse rate' was included. I show that the inclusion of lapse rate allows for the occurrence of stochastic resonance in terms of the performance metric $d'$. At the same time, I show that high levels of lapse rate cause stochastic resonance to disappear. It is also shown that noise generated in the brain (i.e., internal noise) may obscure any effect of stochastic resonance in experimental settings. I further relate the model to a standard equivalent noise model, the linear amplifier model, and show that the lapse rate can function to scale the threshold versus noise (TvN) curve, similar to the efficiency parameter in equivalent noise (EN) models. Therefore, lapse rate provides a psychophysical explanation for reduced efficiency in EN paradigms. Furthermore, I note that ignoring lapse rate may lead to an overestimation of internal noise in equivalent noise paradigms. Overall, describing stochastic resonance in terms of signal detection theory, with the inclusion of lapse rate, may provide valuable new insights into how human performance depends on internal and external noise.
\end{abstract}

\pacs{}

\maketitle

\section{Introduction}
The brain is an inherently noisy system; much of the brain's activity is not driven by external stimulation, or by purposeful internal processes, but by seemingly random activity: noise.  Noise poses a fundamental problem for information processing \cite{VonNeumann1956,SHANNON2001} as it increases variability and limits the clarity of a signal. Yet, given the abundance of noise in neural processing, the brain still achieves remarkably stable perception, presumably because the brain adapted to its own noisiness and that of its inputs. Therefore, studying how the brain responds to noise may help reveal the internal workings of the brain.  

Noise is often considered to limit optimal performance \cite{VonNeumann1956,SHANNON2001}. Indeed the very definition of the performance metric $d'$ is signal-to-noise ratio, where increases in noise decrease $d'$ \cite{GREEN1974signal}. While noise is expected to degrade performance in general, noise can actually improve performance. For example, noise can push a subthreshold signal --- that would normally lead to chance performance in behavioral tasks--- above threshold, and thereby lead to above-change level performance, an effect called stochastic resonance (SR) \cite{McDonnell:2009aa}. Note that SR here refers to any occasion where noise increases performance, and not just cases with periodic input \cite{McDonnell:2009aa}. SR causes optimal performance (i.e., highest detectability) to be reached at non-zero levels of noise. This has potentially important implications, because it suggests that inducing SR by adding noise can be used to boost performance of humans and machines. Because of this potential beneficial effect of noise, it is useful to study SR in more detail, specifically in relation to human performance.

In humans, the influence of noise on performance or perception is often investigated using paradigms in which external noise is added to the signal, and performance thresholds are measured \cite{Lu:2008aa}. A common method is called the equivalent noise (EN) paradigm, which is often used to estimate internal (brain-generated) noise \cite{Lu:2008aa}, and has its origins in engineering to measure noise dependence in electronic amplifiers \cite{north1942absolute}. In the EN paradigm, the overall noise $\sigma$ is assumed to be a combination of internal noise $\sigma_\text{int}$ and externally added noise $\sigma_\text{ext}$, whose variances add: $\sigma^2 = \sigma_{\text{int}}^2 + \sigma_{\text{ext}}^2$. In experimental settings, various amounts of external noise are added to a signal and the effects on detection thresholds are mapped out. When $\sigma_\text{ext} \ll \sigma_\text{int}$, only internal noise limits performance. At high levels of external noise ($\sigma_\text{ext} \gg \sigma_\text{int}$), performance is determined by external noise. At intermediate amounts of noise, the threshold-versus-noise (TvN) curve shows an elbow separating the two regimes, which is located at the level where internal and external noise are equivalent; the location of the elbow thus represents the level of internal noise. 

Although extensively used, the EN paradigm often only considers 2 parameters of interest (i.e., internal noise, and efficiency). Although several extensions on the basic EN paradigm exist \cite{Lu:2008aa} they almost invariably assume that the brain is optimal at setting decisions thresholds. However, human observers do not always set decision thresholds optimally \cite{GREEN1974signal}. This sub-optimal behavior allows for SR \cite{GONG2002}, which indeed has been reported in human observers \cite{Simonotto:1997:aa,Collins:1996aa,Moss:2004aa,Ward:2002aa,GORIS2008}. The fact that SR occurs indicates that the effects of noise are more complicated than often modelled in EN approaches.

Stochastic resonance has been extensively investigated from an engineering point of view, but less so in humans.  A good description of SR in humans is more complicated because the human sensory system presents some challenges in researching stochastic resonance. First, the noise may originate from one of many processing levels in the brain, from early sensory, to later decisional stages, each potentially having different effects on performance. Second, humans are not machines, and they suffer from attentional lapses (induced by, e.g., decreased arousal or vigilance). The influence of attentional lapses has not been studied in the stochastic resonance literature. To determine the optimal level of external noise to achieve optimal performance (i.e., maximal SR), these two specifically human challenges need to be better understood.

Here, I model the influence of noise in sensory signals and decision criteria. To provide further insight into the usefulness of SR as a measure of the influence of noise on human perception, I will compare this model to the EN paradigm \cite{Lu:2008aa,PELLI1981-thesis}.

\section{A signal detection model of stochastic resonance}
Previous work has shown that signal detection theory can provide a good framework for modeling stochastic resonance in human behaviour. Gong et al. \cite{GONG2002} used the following equations: 
\begin{equation}
H =  \int_{c}^{\infty} \mathcal{N}(s,\sigma^{2}) \label{eq:hitrate}
\end{equation}
\begin{equation}
F = \int_{c}^{\infty} \mathcal{N}(n,\sigma^{2})\label{eq:falsealarmrate}
\end{equation}
\begin{equation}
p = \frac{1}{2} (H + (1-F))
\end{equation} 
where H is the hit rate, and F is the false alarm rate, $c$ is a criterion (i.e., threshold), $s$ is the signal strength, $n$ is the mean noise strength, $\mathcal{N}(.)$ is the normal distribution with a certain mean ($s$ or $n$) and variance ($\sigma^2$).  The variable $\sigma$ represents the noise in the system.  This model produced SR for accuracy ($p$), when $c > s$.

This model also predicted that in the limit of zero noise, $d'$, another measure of (human) performance, increased to infinity. Although this may make sense when $d'$ is interpreted as a signal-to-noise ratio, it does not when $d'$ is considered a metric of human performance. As a measure of human performance, $d'$ should drop to chance level when noise is decreased, just as accuracy does. Therefore, I extended the model presented in \cite{GONG2002}, with another human characteristic, namely lapse rate $\lambda$. 
\begin{equation}
H = \lambda + (1 - 2\lambda) \int_{c}^{\infty} \mathcal{N}(s,\sigma^{2}) \label{eq:hitrate}
\end{equation}
\begin{equation}
F = \lambda + (1 - 2\lambda) \int_{c}^{\infty} \mathcal{N}(n,\sigma^{2})\label{eq:falsealarmrate}
\end{equation}
\begin{equation}
d' = \Phi^{-1}(H)-\Phi^{-1}(F). \label{eq:dprime}
\end{equation}
where  $d'$ is an unbiased measure of performance, $\lambda$ is the lapse rate ($0 \le \lambda \le 1/2$) , $\Phi^{-1}$ is the inverse of the cumulative normal distribution function. The variable $\sigma$ again represents the noise in the system, but for our purposes it is important to realise that this noise can have an external ($\sigma_\text{ext}$) or internal (brain-generated; 
$\sigma_\text{int}$) origin, as $\sigma_\text{ext}$ is under control of the experimenter, while $\sigma_\text{int}$ is an internal property of the system under study, i.e. the human participant. 

The lapse rate $\lambda$ represents the proportion of guesses that a participant is required to make due to not paying attention to the stimulus. When $0 < s < c$  (i.e. sub-threshold) and $\sigma$ is small, H and F approach $\lambda$, and $d'$ approaches 0, as required for a measure of performance. Importantly, without the inclusion of $\lambda$, both H and F approach 0 as $\sigma$ decreases, but F much faster then H, because $n < s $, leading $d'$ to increase drastically (see equation \ref{eq:dprime}).

\begin{figure}
\includegraphics{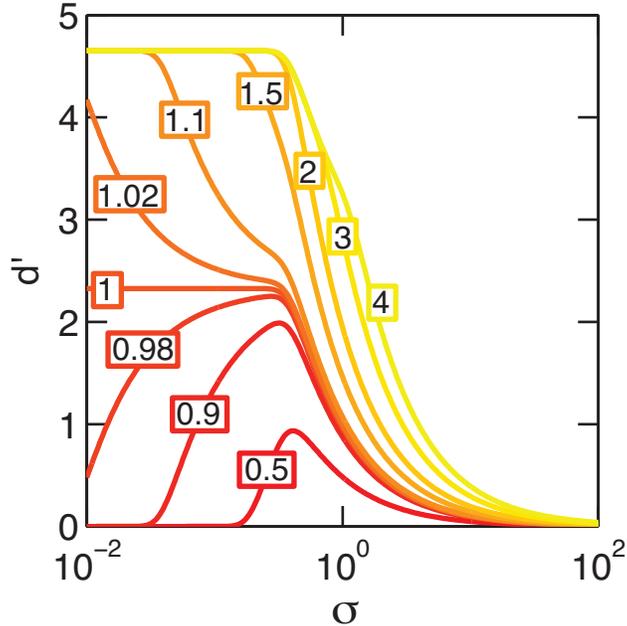} %
\caption{Dependence of $d'$ on noise ($\sigma$), and signal strength ($s$). Parameters are $\lambda = 0.01$, and $c = 1$, while the different lines are solutions at different values of $s$ (indicated with the boxed numbers). \label{fig:SR-dprime}}
\end{figure}

When $\lambda > 0$, the model can produce SR for $d'$, as shown in Fig.\ \ref{fig:SR-dprime} in which the dependence of $d'$ on $\sigma$ is shown for various levels of signal strength $s$ (indicated by the boxed numbers), with $\lambda = 0.01$, and $c = 1$. When signal strength is below criterion $c$ (i.e., $s <  1$) --- which without SR would lead to chance performance --- performance rises above chance (i.e., $d' > 0$) for intermediate levels of $\sigma$. This is a form of stochastic resonance. At small $\sigma$, $d'$ does not increase to infinity (as was found by \cite{GONG2002}), but instead
\begin{equation}
     \lim_{\sigma\to0} d' = 
\begin{cases}
    0,						& \text{if } s < c\\
    -\Phi^{-1}(\lambda),              	& \text{if } s = c\\
    -2 \Phi^{-1}(\lambda) 	       	& \text{if } s > c.\\
\end{cases}
\end{equation}
This behaviour is consistent with human performance, where subthreshold signals (i.e., $s < c$) lead to chance performance, and supra-threshold performance is limited by attentional lapse rate.

\section{Detection thresholds}
In psychophysical experiments, it is common to determine the stimulus strength at which performance reaches some predefined performance threshold at various levels of externally applied noise. Common thresholds in two-choice or 2 alternative forced choice tasks are 75\% accuracy, and a $d'$ of 1. 

By solving eq.\ \ref{eq:dprime} for $s$, one can calculate the threshold signal strength $s_{th}$, and thus TvN curves. Using Mathematica$\textsuperscript{\textregistered}$ 10 it was found that
\begin{subequations}
\begin{align}
s_\text{th}^\text{\tiny{SR}} &= c+\sqrt{2} \text{erf}^{-1}(A)\ \sigma \label{eq:solvedfors}\\ 
		A &=\frac{\text{erf}\Big[
			\frac{d'}
			{\sqrt{2}}- \text{erf}^{-1}\big((1-2\lambda) \ \text{erf}(\frac{c-n}
												 	   {\sqrt{2}\sigma})\big)\Big]}
		{1 - 2\lambda},\label{eq:solvedforB}
\end{align}
\end{subequations}
where $\text{erf}$ and $\text{erf}^{-1}$ are the error function, and inverse error function, respectively. Note that we will use the superscript \textsc{SR} to refer to $s_\text{th}$ specifically in equation \ref{eq:solvedfors}. When referring to the threshold signal strength more generally, no superscript it used. In this paper, we will assume $n = 0$.
%

When $\lambda = 0$, this equation reduces to a straight line
\begin{equation}
s_\text{th} = d'\sigma,\label{eq:sds}
\end{equation}
independent of the criterion, and without a possibility to yield SR. Rewriting eq. \ref{eq:sds} yields $d'=s_\text{th}/\sigma$, which is the signal-to-noise ratio \cite[c.f.,][]{GONG2002}.
\begin{figure}
\includegraphics{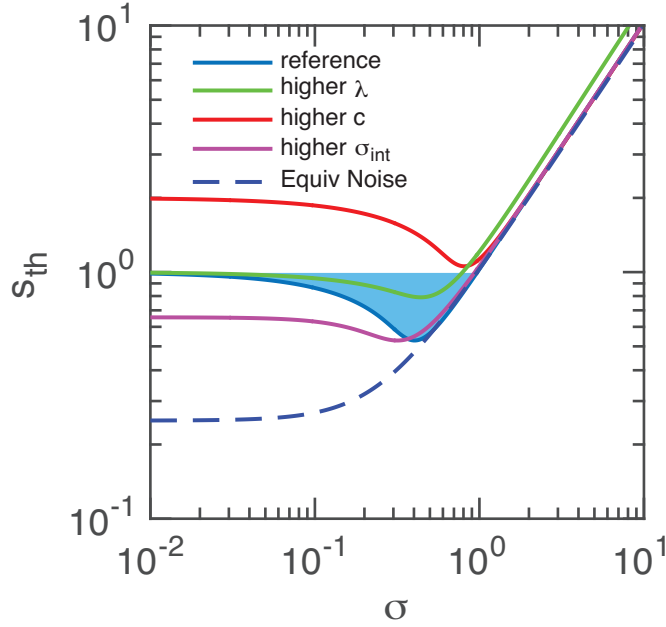} %
\caption{Dependence of detection threshold $s_\text{th}$ on noise $\sigma$. All parameters are as for the reference line (and as in Fig.\ \ref{fig:SR-dprime}), except where otherwise specified. Reference: $d' = 1$, $c = 1$, $\lambda = 0.01$; higher c: $c = 2$; higher $\sigma_\text{int}$: $\sigma_\text{int} = 0.25$; higher $\lambda$: $\lambda = 0.07$; Equivalent Noise: $\sigma_\text{int} = 0.25$, $\lambda = 0$, $c = s/2$. For the `higher $\sigma_\text{int}$', and `equivalent noise' curves, the x-axis represents $\sigma_\text{ext}$.
\label{fig:SR-threshold}}
\end{figure}

When taking $\lambda > 0$, SR is possible.  Equation \ref{eq:solvedfors} is plotted for various parameter combinations in Fig.\ \ref{fig:SR-threshold}. An (arbitrary) base-line curve (blue) describes a characteristic `dipper' function [or threshold-versus-noise (TvN) function], which has been repeatedly reported in the literature for various masking paradigms \cite{Solomon:2009aa}. The curve shows that the detection threshold is lowest at intermediate noise levels $\sigma$, characteristic of SR (highlighted in blue shading). In the following sections, the influence of various factors on the shape of this curve, and the appearance of stochastic resonance are investigated. 

\subsection{The influence of $\lambda$}
First, the influence of lapse rate $\lambda$ is investigated. With an increase in $\lambda$ (green line; $\lambda = 0.07$) relative to the reference curve, SR decreases in magnitude, and the point of maximal SR moves to higher noise levels. The maximum level of $\lambda$ before SR is lost can be calculated by setting eq. \ref{eq:solvedforB} to zero. Solving for $\lambda$ in the limit of zero noise gives the maximum value of $\lambda$ (i.e., $\lambda_{\text{max}}$):
\begin{equation}
\lambda_{\text{max}} = 1-\Phi(d'). \label{eq:lmax}
\end{equation}

\begin{figure}
\includegraphics{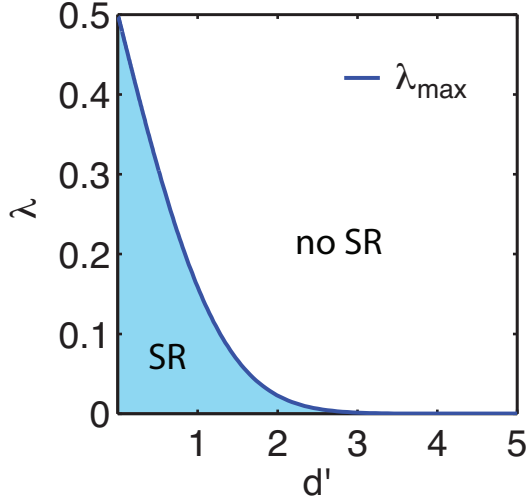} %
\caption{The maximum lapse rate, dependent on $d'$. The shaded area allows for SR.
\label{fig:dprime-lapse}}
\end{figure}

Figure \ref{fig:dprime-lapse} shows the dependence of $\lambda_{\text{max}}$ on $d'$. At $d' = 1$,  a typical value of $d'$ in psychophysical experiments, $\lambda_\text{max} \approx  0.1587$. This level of $\lambda_\text{max}$ is high compared to typical lapse rates around 0.06 or less \cite{Wichmann:2001aa}, and suggests that $\lambda$ is not a limiting factor on the occurrence of SR in many experiments.

A final observation is that the effects of $\lambda$ remain visible at large values of $\sigma$, while other factors, discussed later, do not have a mayor effect at large $\sigma$ (see Fig. \ref{fig:SR-threshold}). 

\subsection{The influence of criterion, and decision noise}
When the decision criterion is set optimally (between $s$ and $n$), SR cannot take place. However, c is often set suboptimally \cite{GREEN1974signal}. The level at which criterion $c$ s set, has a strong influence on the vertical position of the TvN curve.  With a higher $c$, the curve moves upward, and maximal SR moves to higher noise levels (red line, Fig. \ref{fig:SR-threshold}). In a biological system, the setting of this decision boundary $c$ probably depends on the amount of noise in the decision system ($\sigma_\text{d}$). The parameter $c$ would sensibly be set at a level which is high enough to prevent many false positives, but low enough to prevent too many misses. Thus, where $c$ is put determines how liberal or conservative the decision stage is. A `present' response could have the requirement that the decision signal is larger than the $P$th percentile of response distribution in the decision stage in the absence of input (i.e., purely noise-driven activity in the decision stage, here assumed to be normally distributed). Assuming an unbiased (mean = 0) but noisy response originating from the decision stage, the criterion should be set at
\begin{equation}
c =  \sigma_{\text{d}}\Phi^{-1}(P). \label{eq:cdecision}
\end{equation}
Therefore, $c$ can be interpreted as reflecting noise in a post-sensory decision stage. If $\sigma_{\text{d}}$ increases, so does c, which results in an upwards shift in the detection threshold $s_\text{th}$, as can be observed in Fig.\ \ref{fig:SR-threshold}. Conversely, because SR can only occur when $c > s$, systems with low decision noise (or alternatively, very liberal systems with a low $p$) will not show SR. The condition for SR is that
\begin{equation}
\sigma_\text{d} > \frac{s}{\Phi^{-1}(P)}.
\end{equation}

As an aside, in signal detection theory, optimal performance on a task is reached when the criterion is set between mean of the signal and noise distribution (see e.g. \cite{GONG2002,GREEN1974signal}). In our case this means $c = s/2$. Because $c$ is smaller than $s$, no SR will occur (see also \cite{GONG2002}). However, because $c$ is larger than zero, the optimal decision noise $\sigma_{\text{d}}$ is, counterintuitively, larger than zero:
\begin{equation}
\sigma_\text{d} = \frac{s}{2\Phi^{-1}(P)}.
\end{equation}
This conclusion is consistent with a recent analysis on economic decisions making \cite{TSETSOS2016}.

\section{The influence of internal noise, and a comparison to the Equivalent Noise paradigm}
In our model, the noise $\sigma$ is the total amount of noise in the sensory-perceptual system. One can subdivide $\sigma$ into two independent  components: (i) noise that is external to the system $\sigma_\text{ext}$, and e.g. is present in the stimulus; (ii) noise that is internally-generated $\sigma_\text{int}$, where their variances add. In the equivalent noise paradigm, such as the linear amplifier model \cite{Lu:2008aa}, this is often rewritten as:
\begin{equation}
\sigma = {\sqrt{\sigma_\text{int}^2+\sigma_\text{ext}^2}}.
\end{equation}

When inserting this into eq. \ref{eq:solvedfors}, one can show the effect of adding a constant amount of $\sigma_\text{int}$ on performance thresholds (Fig.\ \ref{fig:SR-threshold}; purple line). Increases in $\sigma_\text{int}$ result in left-ward shift of the TvN curve relative to the reference curve. The increase in $\sigma_\text{int}$ simultaneously results in lower thresholds (better performance) at low $\sigma_\text{ext}$, while it increases thresholds at high $\sigma_\text{ext}$. The former result is due to $\sigma_\text{int}$ itself causing SR relative to the condition where $\sigma_\text{int} = 0$. In effect, it is moving the SR dip to the left. With even higher levels of internal noise, the curve is shifted so far to the left that upward arm of the curve at low noise levels, disappears. Even though $\sigma_\text{int}$ still increases performance relative to the reference curve in this case, and thus shows SR, it will not be recognised as such because it does not present itself as a dip in the TvN curve. I come back to this in the discussion. Increasing $\sigma_\text{int}$ even further will remove SR completely.

\subsection{A comparison to the equivalent noise paradigm}
In stead of assuming a fixed criterion $c$, as in the derivations above, the equivalant noise paradigm assumes that the decision criterion is set optimally, i.e, $c = s/2$. When this is inserted into eq. \ref{eq:dprime}, and deriving the threshold performance, we obtain:
\begin{equation}
s_\text{th}^\text{\tiny{EN}} = 2\sqrt{2}\ \text{erf}^{-1}\Bigg(\frac{\text{erf}\big(\frac{d'}{2\sqrt{2}}\big)}{1-2\lambda}\Bigg)\ \sqrt{\sigma_\text{int}^2+\sigma_\text{ext}^2}. 
	\label{eq:th_EN}
\end{equation}
Setting $\lambda = 0$ (one of the assumptions often made) results in the standard dependence of the detection threshold on internal an external noise 
\begin{equation}
s_\text{th}^\text{\tiny{EN}} = {d'\sqrt{\sigma_\text{int}^2+\sigma_\text{ext}^2}}
\end{equation} 
(when $d'$ is set to 1). This dependence is plotted in Fig \ref{fig:SR-threshold} (dashed curve). Due to the fact that $c < s$, there is no SR possible in this description of human performance.

\subsection{The relationship between $\lambda$ and efficiency $\eta$}
Because measured thresholds in psychophysical experiments are often higher than the optimal thresholds, the EN paradigm  often includes an ``efficiency" parameter $\eta$ when fitting the curve to experimental data:
\begin{equation}
s_\text{th} = \frac{d'}{\eta}\sqrt{\sigma_\text{int}^2+\sigma_\text{ext}^2},
\label{eq:ENefficiency}
\end{equation}
where $\eta < 0$, which scales the curve up (i.e., higher detection thresholds).

When reordering Eq.\ \ref{eq:th_EN} one can see that $\lambda$ provides a scaling factor in our model, similar to $\eta$ in equivalent noise paradigms. This relationship makes intuitive sense, as an increased number of lapses, decreases efficiency:  
\begin{equation}
\sigma_\text{th}^\text{\tiny{EN}} =\text{AE} \ \sqrt{\sigma_\text{int}^2+\sigma_\text{ext}^2}.
\end{equation}
where AE is a constant, and related to the calculation efficiency (cf. \cite{PELLI1981-thesis}), but here abbreviated as AE (attentional efficiency) to capture the fact that it is determined by attentional lapse rate $\lambda$:
\begin{equation}
\text{AE} = 2\sqrt{2}\ \text{erf}^{-1}\Bigg(\frac{\text{erf}\big(\frac{d'}{2\sqrt{2}}\big)}{1-2\lambda}\Bigg).
\end{equation} 

\begin{figure}
\includegraphics{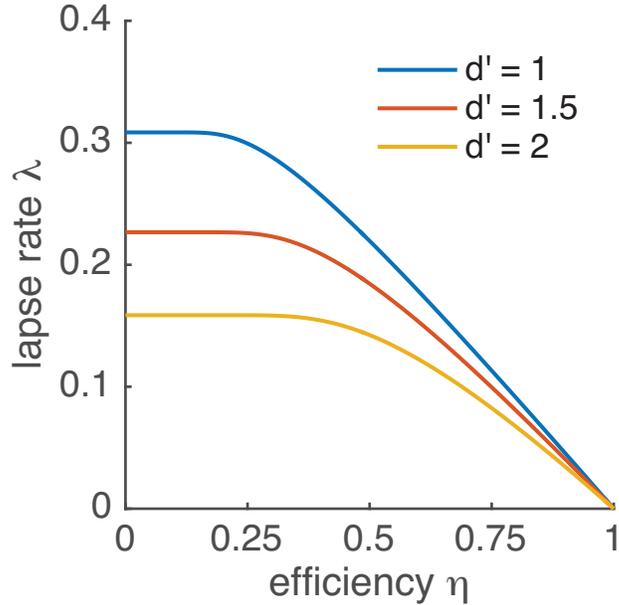} %
\caption{The relationship between efficiency parameter $\eta$ in equivalent noise paradigms, and lapse rate $\lambda$.
\label{fig:efficiency}}
\end{figure}

When equating AE to $d'/\eta$  (see eq. \ref{eq:ENefficiency}) to derive the relationship between $\lambda$ and $\eta$, one finds that
\begin{equation}
\eta = \frac{d'}{2\sqrt{2}\ \text{erf}^{-1}\Big(\frac{\text{erf}(\frac{d'}{2\sqrt{2}})}{1-2\lambda}\Big)},
\end{equation}
and 
\begin{equation}
\lambda = -\frac{\text{erf}\big(\frac{d'}{2\sqrt{2}}\big)-\text{erf}\big(\frac{d'}{2\sqrt{2}\eta}\big)}{2\ \text{erf}\big(\frac{d'}{2\sqrt{2}\eta}\big)} = \frac{\Phi(\frac{d'}{2})-\Phi\big(\frac{d'}{2\eta}\big)}{1-2 \Phi\big(\frac{d'}{2\eta}\big)}.	
\end{equation}
This relationship is plotted in Fig.\ \ref{fig:efficiency} for different values of $d'$. These results show that lower efficiency can be captured by increased lapse rates. 

Overall, the comparison between our model and the equivalent noise paradigm (specifically, the linear amplifier model) suggests that $\eta$ can expressed in terms of $\lambda$, and thus that lapse rate $\lambda$ may explain at least part of the suboptimal efficiency (i.e., $\eta < 1$) that is often reported in psychophysical experiments.

\subsection{The pooling parameter}
In some psychophysical experiments, performance thresholds are actually lower than predicted. This would imply an efficiency $> 1$. Generally, however, the results are interpreted differently, with total noise calculated as:  
\begin{equation}
\sigma = \sqrt{\frac{\sigma_\text{int}^2+\sigma_\text{ext}^2}{n}}, \label{eq:pooling}
\end{equation}
where $n > 1$, representing a pooling parameter, which replaces $\eta$. Equation \ref{eq:pooling}, in essence, describes an important property of the central limit theorem, and $n$ quantifies the number of samples that are taken to estimate a mean. For example, how many individual moving dots are combined to estimate a global pattern motion \citep[e.g.,][]{Dakin:2005aa}. 

Detection thresholds can be obtained from eq.\ \ref{eq:solvedfors} or eq.\ \ref{eq:sds}, by replacing $\sigma$ with $\sqrt{(\sigma_\text{int}^2+\sigma_\text{ext}^2)/n}$, while setting $c = s/2$,  $\lambda = 0$:
\begin{equation}
s_\text{th} = d' \sqrt{\frac{\sigma_\text{int}^2+\sigma_\text{ext}^2}{n}}, \label{eq:clt}
\end{equation}

Equation \ref{eq:clt} and eq. \ref{eq:ENefficiency}, are very closely related as they describe the same relationship when $n$ is equal to $\eta^2$. 

 I set $\lambda = 0$, here, because otherwise both $\eta$ (or $\lambda$) and $n$ need to be determined, and as both have the same scaling effect, the system would be underdetermined. However, in experimental settings, unless participants are experienced and extremely motivated, it is unlikely that $\lambda = 0$, and thus a correct value of the pooling parameter $n$ can only be obtained if $\lambda$ is also determined.

\subsection{Overestimation of internal noise in the EN paradigm}
When comparing the EN results in Figure \ref{fig:SR-threshold} to those of the `high $\sigma_\text{int}$' condition, one can observe that the suboptimal positioning of the criterion $c$ in the latter case caused an expected increased detection thresholds at low external noise values (left side of the plot). An important consequence is that the curves are quite different even though the internal noise is the same. Consequently, fitting the equivalent noise function to experimental data with SR will give incorrect (over)estimates of internal noise.
\begin{figure}
\includegraphics[width=80mm]{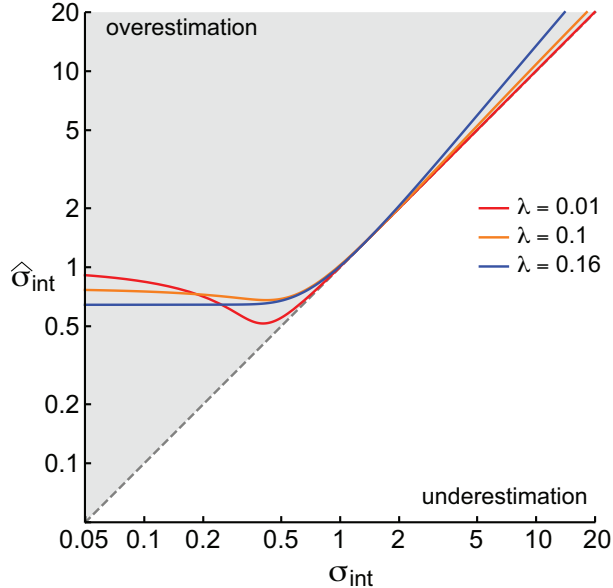}%
\caption{(a) The dependence of $\hat{\sigma}_\text{int}$ on $\sigma_\text{int}$, when SR data is fit with the EN equation. Internal noise is overestimated, especially at small levels of $\sigma_\text{int}$.
\label{fig:intnoiseestim}}
\end{figure}

One can approximate the overestimation of internal noise by equating the EN model (eq. \ref{eq:th_EN}) and the SR model (eq. \ref{eq:solvedfors}) at $\sigma_\text{ext} = 0$. Then, while taking the noise in the SR equation as the actual internal noise ($\sigma_\text{int}$) and that in the EN equation as the estimated internal noise ($\hat{\sigma}_\text{int}$), we obtain:
\begin{equation}
\hat{\sigma}_\text{int} = s_\text{th}^\text{\tiny{SR}}/\text{AE},
\end{equation}
which is plotted in Figure \ref{fig:intnoiseestim}. Estimated internal noise is close to actual internal noise when the actual internal noise is large. However, when the actual internal noise is low, the estimated internal noise is overestimated. When $\lambda$ is relatively low, the estimated internal noise is not even monotonically related to the actual internal noise. Therefore, the EN paradigm overestimates low internal noise. The precise amount of overestimation depends on various parameters (e.g., $\lambda$, $c$, $d'$), and on the distribution of external noise values tested in experimental settings, and how much the dip caused by SR influences the EN model fit.

\section{Stochastic resonance in terms of accuracy}
\begin{figure}
\includegraphics{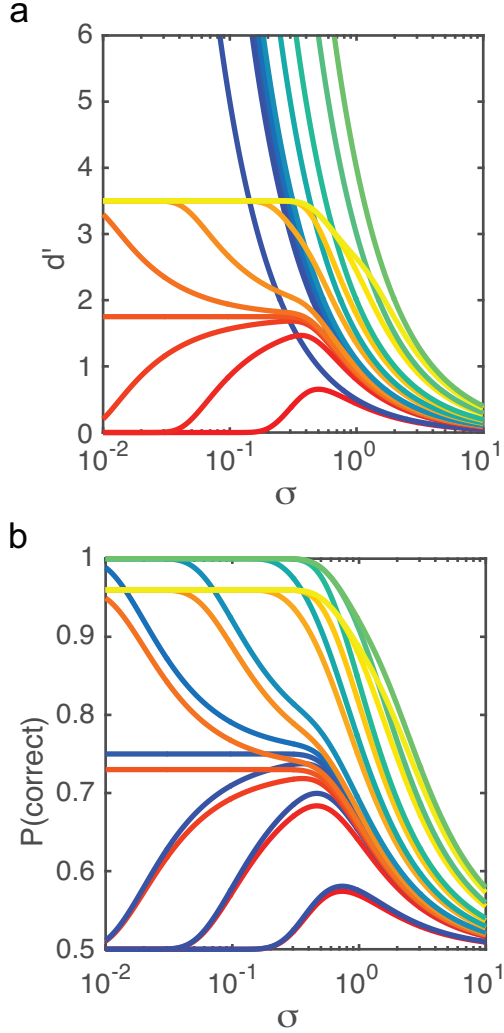} %
\caption{(a) The dependence of $d'$ on noise $\sigma$. The warm colors are for the current model, the cold colors are for those of \cite{GONG2002}. The models are very different at low $\sigma$, but converge at high $\sigma$. (b) The dependence of accuracy [P(correct)] on $\sigma$. The current model is a scaled version of the model by \cite{GONG2002}. Parameter were $c = 1$, $\lambda = 0.04$.
\label{fig:modelcomp}}
\end{figure}
I have focussed so far on the dependence of SR on $d'$, because data in terms of accuracy have previously been reported on in a similar model without $\lambda$ \cite{GONG2002}. A direct comparison between the current model and that of \cite{GONG2002} in terms of $d'$ and accuracy is made in Fig.\ \ref{fig:modelcomp}. As one can see, the largest difference is in terms of $d'$, and not in terms of accuracy.
In fact, the level of above-chance accuracy in our model is $1-2\lambda$ times that of the model where $\lambda = 0$ \cite{GONG2002}. This fixed relationship means the analyses by \cite{GONG2002}, which rested on finding the point where the derivative of accuracy relative to $\sigma$ was 0, will remain valid in the current model. 

\section{Discussion}
I have presented a model that explains SR in a signal detection framework, adding the human characteristic of lapse rate. The inclusion of lapse rate allows for stochastic resonance to be calculated at various levels of the performance metric $d'$, in addition to accuracy. The model produces experimentally observed `dipper' functions, when plotting threshold signal strength versus noise (TvN). I compared the current model of SR to the equivalent noise paradigm, and show that the lapse rate may be used to explain the suboptimal `efficiency' that is often found in experimental paradigms, as it scales the TvN curve up. I also argue that fitting data with the equivalent noise approach when SR is present can lead to incorrect estimates of the level of internal noise. 

In our model the lapse rate has a strong influence on SR. With a typically employed performance threshold of $d' = 1$, the maximum lapse rate is about 0.16 before SR is lost. This is relatively high, and unlikely to be reached in most psychophysical experiments, suggesting that SR could be observed in many experiments. However, lapse rates increase when people perform dual-tasks \cite{Buckley:2016aa}, and when they are tired \cite{Anderson:2012aa}, and thus even with a relatively lenient performance threshold of $d' = 1$, SR may not be observed. When more stringent performance thresholds are employed such as $d' = 2$, or $d' = 3$, the maximum lapse rate rapidly declines to ~0.023, and ~0.0013, which are at or below typically-observed lapse rates \cite{Wichmann:2001aa}, resulting in weak or absent SR.

I also show that the position of the decision criterion has a large influence on the signal thresholds. When criterion is increased, performance thresholds are also much higher. I proposed that the criterion level depends on the noise in the decision stage, as well as on how liberal or conservative the decision stage is. SR is only possible in humans that have a noisy decision stage (large decision noise), or those that are quite conservative (i.e., need a strong signal before they say `present').

The current framework also allows one to model influences of internal noise ($\sigma_\text{int}$). When internal noise is small it allows for SR, but when internal noise is large, SR will disappear, which is a potential explanation of why SR is not observed more often then it is. Low amounts of $\sigma_\text{int}$, cause SR in their own right, just as external noise does. In Fig. \ref{fig:SR-threshold}, this can be observed as a decrease in threshold at low external noise (on the left of the plot), compared to the reference curve. In psychophysical settings, however, this would not be interpreted as $\sigma_\text{int}$-induced SR, because in most experiments internal noise is a fixed value that is not manipulated (and thus a reference curve is lacking). 

One could potentially design experiments to manipulate internal noise. For example, it has been suggested that inattention increases internal noise \cite{Rahnev:2011aa,Lu:1998aa}, which could counterintuitively cause a decrease in detection threshold through SR in inattention conditions. Alternatively, one could use an individual differences approach to investigate whether internal noise causes SR. With this approach one would predict that individuals with moderate levels of internal noise have lower (detection) thresholds than individuals with either low or high levels of internal noise. In this context it is interesting that SR has been proposed as a potential explanation for the increased perceptual functioning of people with autism spectrum disorder (ASD) on some perceptual tasks \cite{Simmons:2009aa}. Specifically, it is argued that people with ASD have increased levels of internal noise \cite{Dinstein:2012aa,Simmons:2009aa,Milne:2011aa}, and thus could outperform typically-developing (TD) individuals ,who have low internal noise, on some tasks due to SR. Incidentally, the TvN curves in Fig. \ref{fig:SR-threshold}, can further explain the general finding that interindividual variation in task performance is larger in the ASD group than in the TD group, because with low noise the TD group would fall on a relatively flat part of the curve, while with intermediate noise, the ASD group would fall in an area of the curve where small changes in internal or external noise can lead to large changes in threshold measurements.


Overall, these speculations suggest that SR may have a more important role in human performance than often realised. Even in the literature, unidentified SR signatures are present in some data  \citep[e.g.,][]{Dakin:2005aa,Mareschal:2008aa,Lu:2006aa}, which were missed because the data were fitted with an equivalent noise approach (which does not allow for SR). Combined with the data that already showed SR \cite{Simonotto:1997:aa,Collins:1996aa,Moss:2004aa,Ward:2002aa,GORIS2008}, these data suggest that beneficial effects of noise (i.e., SR) may be more common than acknowledged, even in human performance. Our model may help determine which factors play important roles in determining when (and in who) SR occurs.

\begin{acknowledgments}
The author thanks Dr. Dror Cohen for suggestions on a previous version of the manuscript.
\end{acknowledgments}

\bibliography{../../libraries/StochasticResonanceBib}

\end{document}